\documentclass[nofootinbib,aps,prl,reprint,superscriptaddress]{revtex4-2}

\usepackage{graphicx}
\usepackage[colorlinks=true,linkcolor=blue,citecolor=blue,urlcolor=blue,bookmarksopen]{hyperref}

\begin{document}

\title{New Upper Bounds on Exotic Neutron Spin-Electron Spin Interactions via Neutron Spin Rotation Measurements in a Compensated Ferrimagnet}

\author{T. Mulkey}
\email[Contact Author: ]{tmulkey1@gsu.edu}
\affiliation{Georgia State University, Atlanta, GA 30303, USA}
\author{K. N. Lopez}
\affiliation{Indiana University/Center for Exploration of Energy and Matter and Indiana University Center for Spacetime Symmetries, 2401 Milo B. Sampson Lane, Bloomington, IN 47408, USA}
\author{C. D. Hughes}
\affiliation{Indiana University/Center for Exploration of Energy and Matter and Indiana University Center for Spacetime Symmetries, 2401 Milo B. Sampson Lane, Bloomington, IN 47408, USA}
\author{B. Hill}
\affiliation{University of Illinois, Urbana, IL 61801-3003, USA}
\author{M. Van Meter}
\affiliation{Indiana University/Center for Exploration of Energy and Matter and Indiana University Center for Spacetime Symmetries, 2401 Milo B. Sampson Lane, Bloomington, IN 47408, USA}
\author{H. Wijeratne}
\affiliation{Georgia State University, Atlanta, GA 30303, USA}
\author{J. C. Long}
\affiliation{University of Illinois, Urbana, IL 61801-3003, USA}
\author{M. Sarsour}
\affiliation{Georgia State University, Atlanta, GA 30303, USA}
\author{W. M. Snow}
\affiliation{Indiana University/Center for Exploration of Energy and Matter and Indiana University Center for Spacetime Symmetries, 2401 Milo B. Sampson Lane, Bloomington, IN 47408, USA}
\author{K. Li}
\affiliation{Indiana University/Center for Exploration of Energy and Matter and Indiana University Center for Spacetime Symmetries, 2401 Milo B. Sampson Lane, Bloomington, IN 47408, USA}
\author{R. Parajuli}
\affiliation{Georgia State University, Atlanta, GA 30303, USA}
\author{S. Samiei}
\affiliation{Indiana University/Center for Exploration of Energy and Matter and Indiana University Center for Spacetime Symmetries, 2401 Milo B. Sampson Lane, Bloomington, IN 47408, USA}
\author{D. V. Baxter}
\affiliation{Indiana University/Center for Exploration of Energy and Matter, 2401 Milo B. Sampson Lane, Bloomington, IN 47408, USA}
\author{M. Luxnat}
\affiliation{Indiana University/Center for Exploration of Energy and Matter and Indiana University Center for Spacetime Symmetries, 2401 Milo B. Sampson Lane, Bloomington, IN 47408, USA}
\author{Y. Zhang}
\affiliation{Oak Ridge National Laboratory, Oak Ridge, TN 37830, USA}
\author{C. Jiang}
\affiliation{Oak Ridge National Laboratory, Oak Ridge, TN 37830, USA}
\author{E. Stringfellow}
\affiliation{Oak Ridge National Laboratory, Oak Ridge, TN 37830, USA}
\author{J. Torres}
\affiliation{Oak Ridge National Laboratory, Oak Ridge, TN 37830, USA}
\author{R. Hobbs}
\affiliation{Oak Ridge National Laboratory, Oak Ridge, TN 37830, USA}

\date{August 19, 2025}

\begin{abstract}
We report a search for exotic spin-spin interactions between neutrons and electrons which could signal new physics beyond the Standard Model using slow neutron polarimetric imaging through a dense medium of polarized electrons. Our dense polarized electron target is a ferrimagnet held at its magnetic compensation temperature, which realizes a polarized electron ensemble with zero net magnetization. We sought the spin rotation of transversely polarized neutrons from a neutron spin-electron spin interaction of the form $V_2=-g_A^eg_A^n\frac{\hbar c}{4\pi}\vec\sigma_e\cdot\vec\sigma_n\frac{e^{-r/\lambda_c}}{r}$, where $g_{A}^{e}$ and $g_{A}^{n}$ are the electron and neutron axial couplings, $\vec{\sigma_e}$ and $\vec{\sigma_n}$ are the electron and neutron spin, and $\lambda_c$ is the interaction range for an exotic axial vector interaction from massive spin-1 boson exchange of mass $\hbar c/\lambda_c$. The resulting average neutron spin rotation angle per unit length, $\frac{d\bar{\phi}_{F5}}{dz}=[0.41\pm6.30\ (stat.)\pm4.4\ (sys.)]\times10^{-3}$ rad/m, is consistent with zero. Our novel approach improves the previous upper limits on the coupling constant product $g_A^eg_A^n$ by several orders of magnitude in the poorly explored $10^{-8}\leq\lambda_c\leq10^{-2}$ range.
\end{abstract}

\maketitle

\textit{Introduction}---Many theories beyond the Standard Model of particle physics predict the existence of weakly-coupled exotic spin-dependent interactions between Standard Model particles~\cite{Leitner64,Weinberg72,Moody84} mediated by spin-0 or spin-1 boson exchange. These interactions can arise from various physical mechanisms, such as the presence of pseudo-Goldstone bosons resulting from incompletely broken symmetries~\cite{pgoldstone1}, compactifications in string theory~\cite{string1,string2,string3}, and other theoretical frameworks~\cite{paraphoton1,paraphoton2,majorons1,zboson1,zboson2,s1boson1,s1boson2,s1boson3,s1boson4,s1boson5}. For experimental searches that probe these interactions within the framework of quantum field theory, the only internally consistent approach to fundamental interactions of point particles that respects both relativity and quantum mechanics, the possible forms of such interactions are highly constrained. The number of free parameters is limited to the boson masses and the various possible coupling constants: scalar and pseudoscalar couplings for spin-0 boson exchange, and vector and axial-vector couplings for spin-1 boson exchange. Since the strength of these interactions is quite weak,  the amplitude from the exchange of a single boson between Standard Model fermions is sufficient to describe the interaction and confront experimental data. For the large subset of experiments which employ measurements among nonrelativistic nucleons and electrons, one can write down a complete set of nonrelativistic potentials involving the spins, momenta, interaction range, and possible couplings of the particles~\cite{dobrescu_spin-dependent_2006,fadeev_revisiting_2019}. Various reviews~\cite{Adelberger_rev,Antoniadis_rev,Safronova_rev} including a very recent work~\cite{Cong_rev} describe the broad array of experimental searches sensitive to various couplings, interaction ranges, and potentials performed to date.

Experimental searches for exotic spin-spin interactions are especially challenging to perform. The magnetic moments of nucleons and electrons in a macroscopic spin-polarized medium normally generate large magnetic fields, and the shorter the range of the exotic interaction, the more difficult it is to suppress the associated magnetic systematic errors using magnetic shielding. Slow neutrons possess a unique combination of properties well suited to experimental searches for possible exotic spin-dependent interactions with ranges between the millimeter and Angstrom scale. The small electric polarizability of neutrons compared to atoms and the ability of slow neutrons to penetrate macroscopic amounts of dense matter with negligible attenuation and decoherence enables one to realize spin-dependent interferometric measurements. Slow neutron beams can be polarized, analyzed, and detected with high efficiency. For all of these reasons, polarized slow neutrons have set the most sensitive limits on many types of weakly coupled spin-dependent interactions at sub-mm scales~\cite{Piegsa2012,Yan13,Lehnert17, Haddock2018b}.

In this Letter, we report stringent new experimental constraints on the axial coupling constant product $g_A^eg_A^n$ over a broad set of interaction ranges $\lambda_c$ from spin-1 boson exchange using polarized slow neutron spin rotation through an ensemble of polarized electrons. Our experiment is to our knowledge the first laboratory experiment to constrain $g_A^eg_A^n$ for interaction ranges between $10^{-8}$ m and $10^{-2}$ m. In the nonrelativistic limit the interaction potential from this exchange can be expressed in the following form~\cite{Cong_rev}:
\begin{equation}
    \label{v2}
    V_2=-g_A^eg_A^n\frac{\hbar c}{4\pi}\vec\sigma_e\cdot\vec\sigma_n\frac{e^{-r/\lambda_c}}{r},
\end{equation} 
where $\vec{s}=\vec{\sigma}/2$ is the spin of the polarized particle and $r$ is the distance between the two interacting particles. For an electron target polarized along the neutron momentum and a neutron beam polarized transverse to the neutron momentum, $V_2$ causes a rotation of the neutron spin about its momentum~\cite{sears} with a rotary power of 
\begin{equation}
\label{nOptical}
    \frac{d\phi_{F5}}{dz}=\lambda_nP_{e}N\Delta b,
\end{equation}
 where $\lambda_n$ is the neutron wavelength, $P_e$ is the number of polarized electrons per molecule, $N$ is the number density of the material, and $\Delta b$ is the difference in scattering amplitudes for the two different relative orientations of the electron and neutron spins. $\Delta b$ can be calculated by applying the Born approximation to $V_2$. Combining Eqns.~(\ref{v2}) and~(\ref{nOptical}) results in a relation between the parameters of the potential and the exotic neutron spin rotation angle 
 \begin{equation}
    \label{constraint}
    \frac{d\phi_{F5}}{dz} = g_A^eg_A^n\frac{m_nc}{\pi\hbar}\lambda_c^2\lambda_nP_eN. 
\end{equation}

To develop a macroscopic polarized electron target for this work which does not also generate a large magnetic field, we exploited the well-known features of ferrimagnetism~\cite{ferri1,ferri2}. Ferrimagnets possess two sublattices with atoms of different magnetic moments and an antiferromagnetic coupling. Normally a ferrimagnet retains a net magnetic moment despite this antiferromagnetic coupling as the magnetic moments of the different atoms are unequal. However, in many cases, the differing temperature dependence of the sublattice magnetizations yields a compensation temperature $T_c$ where the net magnetic moment vanishes. Since the different atoms in each sublattice possess different gyromagnetic ratios and different ratios of the orbital and spin components of the total angular momentum, a non-zero electron polarization persists at $T_{c}$. The recent experiments we have conducted to characterize the internal magnetization of the ferrimagnetic sample material used in this work at and near $T_c$ using SQUID magnetometry, X-ray diffraction, polarized neutron imaging, and neutron spin echo spectroscopy all show that our compensated ferrimagnet is an ensemble of polarized electrons with no net internal or external magnetic field~\cite{hughes_polarized_2025}.

\begin{figure*}
\includegraphics[width=\linewidth]{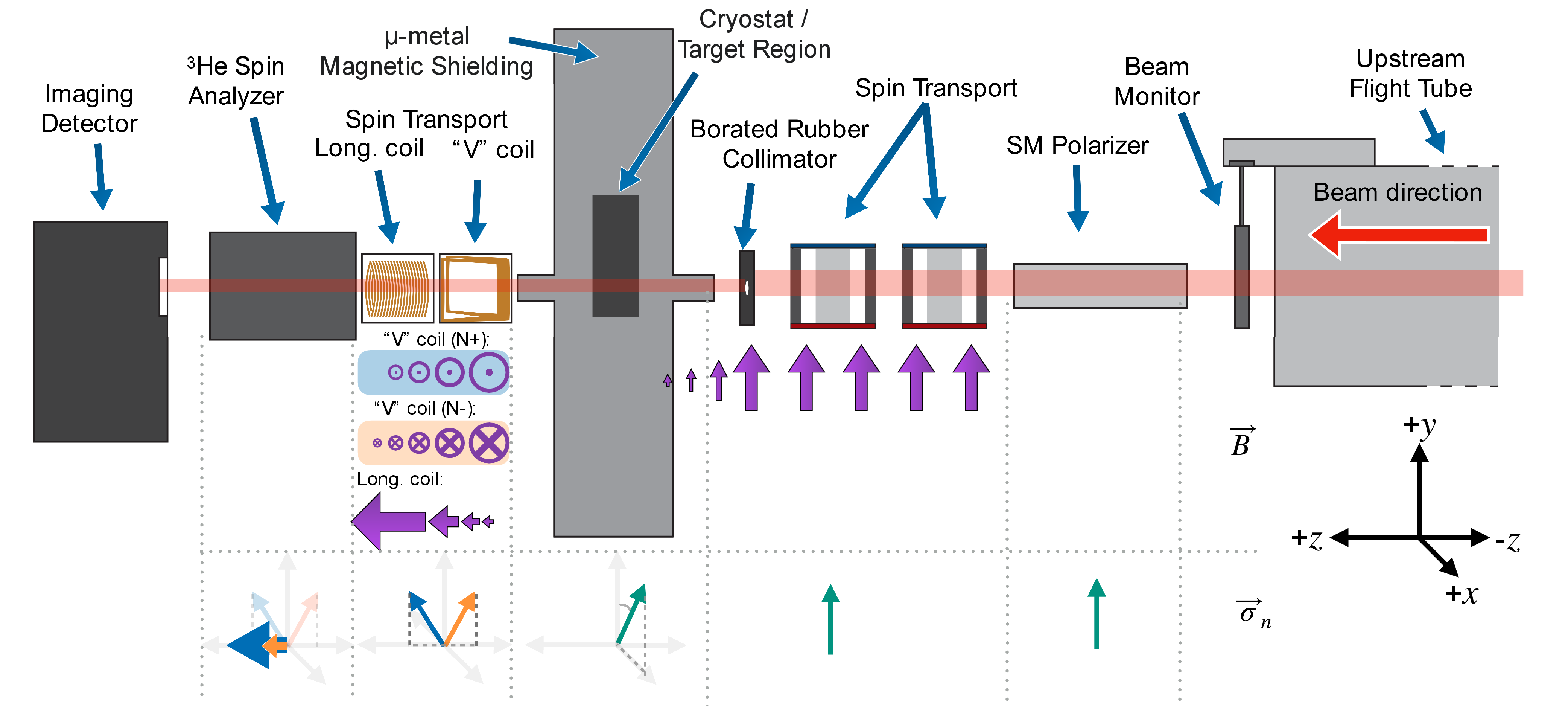}
\caption{\label{apparatus}Slow neutron polarimetry apparatus installed at the CG-1D beamline to measure rotations of the neutron spin into the transverse plane \cite{hughes_polarized_2025}. Magnetic fields for neutron spin transport and effect on the neutron spin for $N_+$ and $N_-$ images are represented below the respective apparatus element.}
\end{figure*}

\textit{Experimental technique, measurements, and results}---Two experiments to measure spin rotation of polarized neutrons in ferrimagnetic terbium iron garnet (TbIG) were conducted at the Oak Ridge High Flux Isotope Reactor CG-1D Multimodal Advanced Radiography Station neutron beamline. The apparatus shown in Fig.~\ref{apparatus} realizes the neutron equivalent of a crossed polarizer-analyzer arrangement well known from visible light optics. A V-cavity neutron polarizer provides a highly polarized slow neutron beam. Magnetic guide fields adiabatically maintain neutron polarization up to the magnetically shielded sample region. The spin transport coils preceding the spin analyzer produce an initial sharp nonadiabatic transition from the low field region and then adiabatically rotate the transverse component of the neutron spin to be parallel or anti-parallel with the neutron polarization analyzer. This relative analyzing orientation is controlled by the direction of the current in this 'V' coil. The analyzer is a longitudinally polarized $^{3}\text{He}$ SEOP neutron spin filter continuously pumped \textit{in situ}~\cite{SEOP1,SEOP2}.

Neutron transmission measurements taken with opposing analyzing orientations generated by reversing the V-coil current determine the neutron spin rotation angle $\phi$ as
\begin{equation}
    \label{An}PAsin(\phi)=\frac{N_+-N_-}{N_++N_-},
\end{equation} 
where $P$ and $A$ are the neutron polarizing and analyzing powers, respectively, and $N_\pm$ is the number of detected neutrons for a given analyzing orientation. 

A preliminary measurement conducted in 2023 mounted the sample in an aluminum cell thermally coupled to a continuous-flow liquid nitrogen cryostat. Mechanical and thermal contact with the faces of the sample were made using sapphire plates, which have low neutron absorption and scattering. The cryostat windows were also sapphire. Silicon diode thermometers placed at various positions on the sample mount enabled temperature control using a PID feedback loop and a resistive heater. A 3-axis fluxgate magnetometer was mounted inside the magnetic shielding close to the cryostat. This apparatus is described in detail in Ref.~\cite{hughes_polarized_2025}.

For the experiment in 2024, the results of which are presented in this paper, the liquid nitrogen cooling system was replaced with a recirculating chiller, which allowed for more stable temperature control. More sensitive fluxgate magnetometer probes placed in closer proximity to the sample delivered more detailed information on changes to the magnetic field near the sample. An additional layer of magnetic shielding further reduced the ambient magnetic field and its fluctuations. The initial experiment used a CCD camera with $2048\times2048$ $42\ \mu m$ pixels and a RC Tritec $^{6}$LiF/ZNS:Cu (ratio 1/2) scintillator; the follow-up experiment used a CMOS camera with $6400\times6400$ $15\ \mu m$ pixels and a Gadox scintillator. Different TbIG samples of purities 88.7\% and 93\% were used for the 2023 and 2024 experiments, respectively. Additional details regarding fabrication, polarization, and purity measurement procedures can be found in Ref.~\cite{hughes_polarized_2025}.

\begin{figure}
\includegraphics[width=1\linewidth]{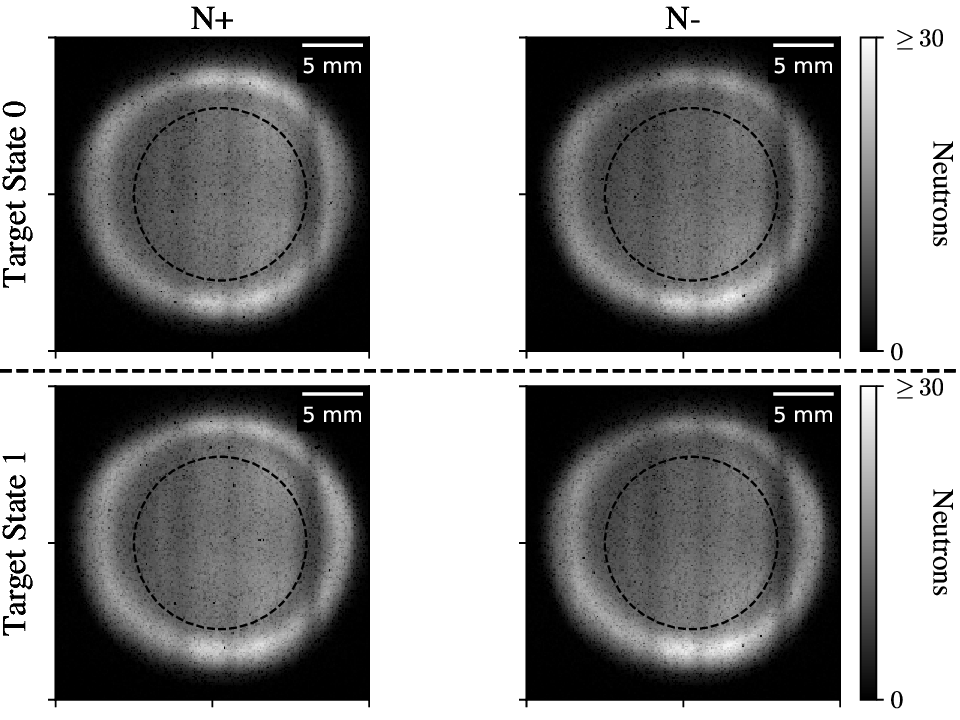}
\caption{\label{images}High contrast $N_{\pm}$ images from the 2024 experiment for each target state. Each image has been normalized to the neutron flux, filtered for hot pixels, and background subtracted. The ROI defining the boundary of the area integrated in determining $N_{\pm}$ is marked by the dashed circle. Nonuniformities in the image intensity come from upstream.}
\end{figure}

Neutron counts are registered in a neutron transmission imaging detector, as shown in Fig.~\ref{images}, of the TbIG samples at $T_c$. Each image is filtered to remove unphysical \lq\lq hot\rq\rq \ pixel artifacts with intensities $5\sigma$ greater than the local average and is background subtracted to remove an underlying pixel value offset. These procedures remove the non-neutron detector backgrounds. The total neutron count $N_{\pm}$ is obtained by summing the image grayscale pixel intensities in a defined Region of Interest (ROI) and dividing by a scaling factor, the average increase in pixel intensity per neutron incident on the detector. The scaling factor was measured independently for each experiment using a neutron beam monitor. The neutron analyzing orientation is flipped after each image and the spin rotation angles are calculated according to Eq.~(\ref{An}). The electron polarization direction is periodically switched between parallel and antiparallel to the neutron momentum through mechanical rotation of the sample. The total neutron spin rotation angle for the two target states labeled 0 and 1, including rotations due to Larmor precession from residual longitudinal background magnetic fields $\phi_B$, can be expressed as $\phi_{0,1}=\phi_{B}\pm\phi_{F5}$.

\begin{figure}
\includegraphics[width=1\linewidth]{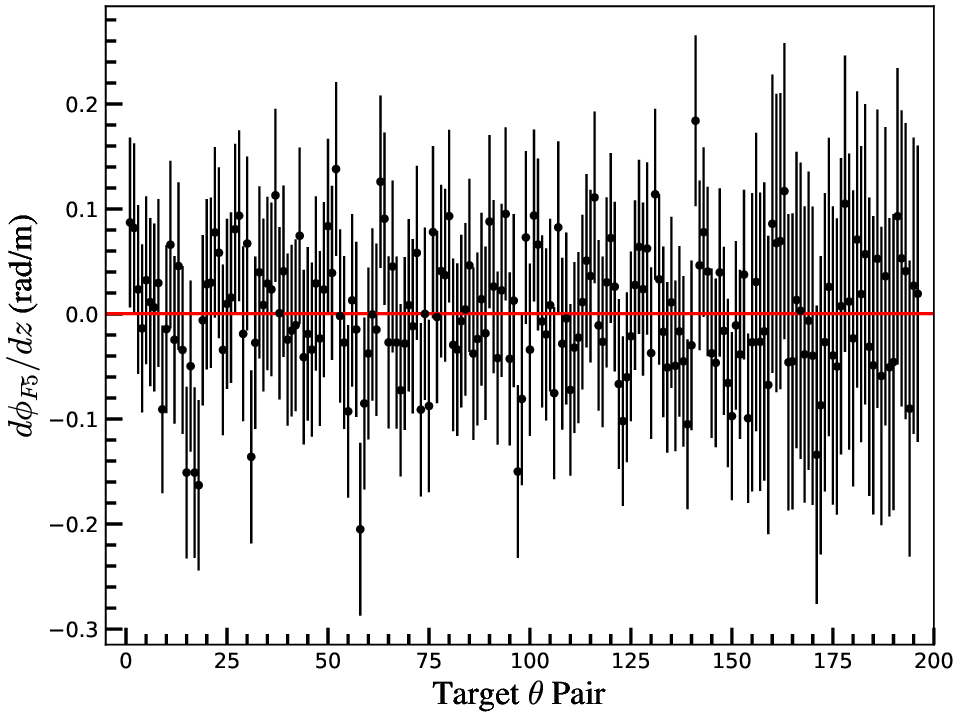}
\caption{\label{thetaF5Fig}(Color online) Measurements of $d\phi_{F5}/{dz}$ from the 2024 experiment. The weighted average rotary power is shown with a red (solid) line $d\bar{\phi}^{2024}_{F5}/{dz}=[0.41\pm6.30\ (stat.)]\times10^{-3}$ rad/m.}
\end{figure}

It is critical to isolate any exotic interaction component $\phi_{F5}$ of the measured neutron spin rotation angle from underlying effects associated with background magnetic fields. Time-dependent drifts in the background magnetic field on the target reversal timescale can generate systematic errors. No time-dependent drifts were resolved during sample rotations in the 2023 experiment. The improved magnetometry of the second experiment enabled more sensitive measurements of the magnetic field at the location of the ferrimagnetic sample. $\phi_B$ can be estimated for each measurement of the total neutron spin rotation angle $\phi_{0,1}$ with a normalization of the average measured magnetic field with the average neutron count asymmetry for each target state pair. By considering the average asymmetry and average magnetic field reading across a pair of target states, the correction factor is insensitive to the absolute accuracy of the magnetic field measurement. Furthermore, any contribution to the measured asymmetry that is coupled to the target state, such as $\phi_{F5}$, is not affected by the correction. To account for any further target state independent variation, $\phi_{F5}$ is calculated as $\frac{\phi_0-\phi_1}{2}$. The rotary power can then be calculated by normalizing to the thickness of the target. The target thicknesses were $L_{2023}=0.74$ cm in 2023 and $L_{2024}=1.0$ cm in 2024. The 2024 results, shown in Fig.~\ref{thetaF5Fig}, were verified by an independent analysis. A blinded analysis procedure accurately recovered an artificial $\phi_{F5}$ injected into the neutron count data.

Systematic effects and their estimated magnitudes are presented in Table~\ref{sysTable}. Due to apparatus design, a small vertical temperature gradient may develop across the sample which can lead to a systematic contribution to the rotary power in the presence of neutron beam intensity nonuniformities and neutron beam centroid offsets from target center. An upper bound on the size of this systematic was determined by splitting the ROI into upper and lower halves and measuring the difference in $d\phi_{F5}/dz$ between each half coupled with our detailed information on the neutron beam intensity distribution from the images. 

Similarly, fluctuations in the sample temperature $T$ can lead to systematic contributions to $d\phi_{F5}/dz$ from the small nonlinearity of the sample magnetization as a function of $ T-T_c$. An upper bound on this systematic was determined from the quadrature difference between the statistical error of the mean measured rotary power and the observed variation of the data $\sigma/\sqrt{N}$.

A traditionally dominant systematic effect in exotic force measurements is the variation in the magnetic field between measurements of $\phi_{0,1}$. The 2024 experiment, which collected data for longer intervals before rotating the target, is more susceptible to this magnetic field drift. This effect is eliminated by the correction procedure which subtracts background asymmetries extracted from the fluxgate probe signals. The variation in the fluxgate signal leads to a 5\% systematic uncertainty associated with the background asymmetries.

The ROI of the images for each experiment was chosen to maximize neutron counts while excluding nonuniform regions near the edge of the sample. Reducing the ROI radius by 10\% leaves $d\phi_{F5}/dz$ unchanged within the slightly higher statistical error. A variation in the neutron beam flux between images can lead to a systematic contribution to the neutron count asymmetry that is independent of any neutron spin rotation. By examining the ratio of neutron beam monitor counts for consecutive images $\frac{BM_{+}}{BM_{-}}$ deviations from unity were found to be less than 1.5\%. No change in $d\phi_{F5}/dz$ within the statistical error was seen upon varying the hot pixel filter rejection criterion from $3\sigma$ to $7\sigma$.

If the temperature of the target deviates slightly from $T_{c}$, nonforward neutron small angle paramagnetic spin flip scattering accepted by the detector can generate a systematic error. Our previous measurements of the intensity of this scattering using neutron spin echo spectroscopy combined with our thermometry precision imply that this systematic is well below $10^{-6}$ rad/m in our apparatus.

\begin{table}
\caption{\label{sysTable}Systematic effects and their errors on $d\phi_{F5}/dz$ in the 2024 experiment. The total systematic error is less than the statistical error.}
\begin{ruledtabular}
\begin{tabular}{l r}
 Source & Uncertainty (rad/m)\\ 
 \hline
 B Variation & $2.1\times10^{-5}$\\
 Flux Variation &  $6.2\times10^{-6}$\\ 
 Filtering Variation & $< 4.1\times10^{-6}$\\ 
 ROI Variation & $< 4.1\times10^{-6}$\\ 
 Temp. Gradient & $2.4\times10^{-3}$\\ 
 Temp. Stability & $3.7\times10^{-3}$\\ 
 SANS & $<1\times10^{-6}$ \\
 \hline
 Total & $< 4.4\times10^{-3}$\\ 

\end{tabular}
\end{ruledtabular}
\end{table}

The weighted average of the measured rotary power $d\bar{\phi}_{F5}/{dz}$ from the 2024 experiment is $[0.41\pm6.30\ (stat.)\pm4.4\ (sys.)]\times10^{-3}$ rad/m, as shown in Fig. \ref{thetaF5Fig}. A 95\% C.L. upper bound on $g_A^eg_A^n$ can be calculated from this result and Eq.~(\ref{constraint}) for interaction ranges up to the size of the sample. The magnitude of the electron polarization for each experiment is estimated to be $P_{e,2023}=0.09\pm0.03$ and $P_{e,2024}=0.3\pm0.09$~\cite{hughes_polarized_2025}. The CG-1D beamline has an average neutron wavelength of $\lambda_n=2.6$~\AA ~\cite{iverson_flux_2024}.

\begin{figure}
\includegraphics[width=1\linewidth]{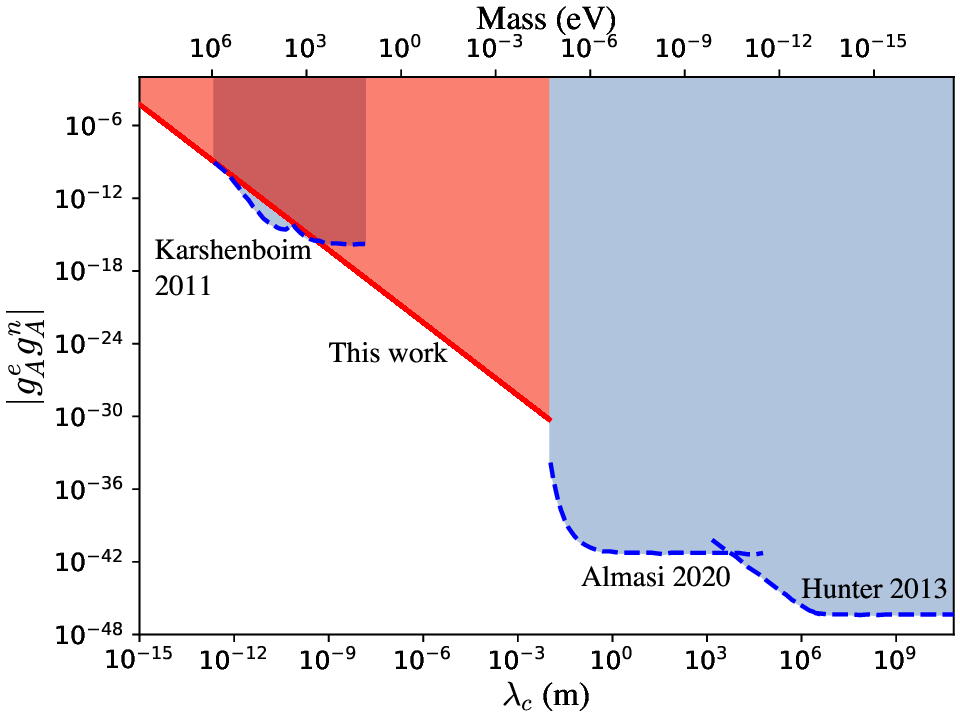}%
\caption{\label{v2Fig}(Color online) Upper limits on the product $g_A^eg_A^n$ as a function of both the interaction range $\lambda_c$ and the boson mass $m_X$ determined from the potential $V_2$. Previous experimental constraints~\cite{karshenboim_hyperfine_2011,almasi_new_2020,hunter_geoelectrons_2013} are noted with blue (dashed) lines. New constraints derived from this work are noted with a red (solid) line. The red (dark) and blue (light) shaded regions denote excluded combinations of $\lambda_c$ and $g_A^eg_A^n$ from this and previous works, respectively.}
\end{figure}

The 95\% C.L. constraint from the 2024 experiment, shown in red (solid line) in Fig.~\ref{v2Fig}, ranges from $g_A^eg_A^n \leq 5.4\times10^{-5}$ at $\lambda_c=10^{-15}$ m to $g_A^eg_A^n \leq 5.4\times10^{-31}$ at $\lambda_c=10^{-2}$ m. The corresponding range of vector boson masses is from $2.0\times10^{8}$ eV to $2.0\times10^{-5}$ eV. This upper bound is the first reported constraint on $g_A^eg_A^n$ in the $10^{-8}$ m to $10^{-2}$ m range derived from $V_2$, bridging the gap between previous constraints measured by atomic hyperfine spectroscopy~\cite{karshenboim_hyperfine_2011} and comagnetometry~\cite{almasi_new_2020}. The red region in Fig.~\ref{v2Fig} shows the combinations of interaction range and coupling magnitude eliminated by this work. 

\textit{Conclusion}---We report a polarized neutron imaging measurement of the rotation angle of neutrons transmitted through a compensated ferrimagnet, from which we derive stringent new upper bounds on $g_A^eg_A^n$ in the $10^{-8}$ m to $10^{-2}$ m range where no sensitive measurements had yet been performed. Our work demonstrates for the first time the power of properly-prepared ferrimagnetic materials and neutron spin rotation techniques to search for exotic spin-dependent interactions between polarized neutrons and polarized electrons over mesoscopic distance scales.  

Our estimated systematic effects are lower than the statistical error, and there is ample room for future improvement of the measurement precision with corresponding improvements to the apparatus to further suppress systematic effects. We envision that the sensitivity of this method can be improved by at least an order of magnitude over the same interaction range by using more intense neutron beams, longer running time, better magnetic shielding, improved magnetometry and temperature control, and improved temperature uniformity to better define and control $T_{c}$. A different ferrimagnetic material with a higher electron polarization density could be developed. Additional exotic spin-spin interactions between neutrons and electrons proportional to $\vec{s_{n}} \cdot (\vec{p_{n}} \times \vec{s_{e}})$ can be sought using the same experimental method by orienting the electron polarization in the target at right angles to both the neutron polarization and the neutron beam momentum. Constraints on additional neutron spin-electron spin interactions will be presented in a subsequent paper. Our data can also be analyzed to constrain the possible existence of the long-speculated manifestation of spin in spacetime geometry known as gravitational torsion~\cite{torsionYuri,torsion1,torsion2,torsion3,torsion4,torsion5,torsion6,torsion7,torsion8}.

\begin{acknowledgments}
\textit{Acknowledgments}---M. Sarsour and T. Mulkey acknowledge support from US Department of Energy grant DE-SC0010443. This research was supported in part by an appointment to the Oak Ridge National Laboratory GRO Program, sponsored by the U.S. Department of Energy and administered by the Oak Ridge Institute for Science and Education. K. N. Lopez, M. Luxnat, M. Van Meter, and W. M. Snow acknowledge support from US National Science Foundation (NSF) grants PHY-1913789 and PHY-2209481 and the Indiana University Center for Spacetime Symmetries. K. N. Lopez also acknowledges the support of the GEM Consortium and Indiana Space Grant Consortium. J. C. Long and B. Hill acknowledge support from grant NSF PHY-1707986. The authors acknowledge the use of facilities supported by NSF through the University of Illinois Materials Research Science and Engineering Center DMR-2309037. This research used resources at the High Flux Isotope Reactor, a DOE Office of Science User Facility operated by the Oak Ridge National Laboratory. The beam time was allocated to MARS on proposal number IPTS-30635 and IPTS-32059. This work was supported by the U.S. Department of Energy (DOE), Office of Science, Office of Basic Energy Sciences, Early Career Research Program Award KC0402010, under Contract DE-AC05-00OR22725.
\end{acknowledgments}

\bibliography{mainbib}

\end{document}